\documentclass[%
 reprint,%
 amssymb, amsmath,%
 aip,cha,%
]{revtex4-1}

\usepackage{docs}%
\usepackage{graphics} 
\usepackage{graphicx}
\usepackage{bm}%
\usepackage[colorlinks=true,linkcolor=blue]{hyperref}%
\expandafter\ifx\csname package@font\endcsname\relax\else
 \expandafter\expandafter
 \expandafter\usepackage
 \expandafter\expandafter
 \expandafter{\csname package@font\endcsname}%
\fi
\begin{document}

\title{Vortex Solution of the Gravitational Field Equation of a Twisted Skyrme Strings}%

\author{Malcolm Anderson$^1$,~Miftachul Hadi$^{1,2}$,~Andri Husein$^3$}%
\email{itpm.id@gmail.com (Miftachul Hadi)}
\affiliation{$^1$Department of Mathematics, Universiti Brunei Darussalam, Negara Brunei Darussalam\\
             $^2$Physics Research Centre, Indonesian Insitute of Sciences, Puspiptek, Serpong, Indonesia\\
		     $^3$Department of Physics, University of Sebelas Maret, Surakarta, Indonesia}%

\begin{abstract}
We construct non-linear sigma model plus Skyrme term (Skyrme model) with a twist in the gravitational field. To simplify the solution, first we examine non-linear sigma model without Skyrme term, in particular with a twist, which comprises a vortex solution with an added dependence on a twist term $mkz$, where $z$ is the vertical coordinate. We find that vortex solution for non-linear sigma model with a twist is similar with vortex solution without a twist. The work is still progress.
\end{abstract}

\maketitle

\section{Non-Linear Sigma Model}
A non-linear sigma model is an $N$-component scalar field theory in which the fields are functions defining a mapping from the space-time to a target manifold \cite{Zakrzewski}. 
By a non-linear sigma model, we mean a field theory with the following properties \cite{hans02}:
\begin{itemize}
\item[(1)] The fields, $\phi(x)$, of the model are subject to nonlinear constraints at all points $x\in\mathcal{M}_0$, where $\mathcal{M}_0$ is the source (base) manifold, i.e. a spatial submanifold of the (2+1) or (3+1)-dimensional space-time manifold.
\item[(2)] The constraints and the Lagrangian density are invariant under the action of a global (space-independent) symmetry group, $G$, on $\phi(x)$.
\end{itemize}

The Lagrangian density of a free (without potential) nonlinear sigma model on a Minkowski background space-time is defined to be \cite{chen}
\begin{equation}\label{1}
\mathcal{L}=\frac{1}{2\lambda^2}~\gamma_{AB}(\phi)~\eta^{\mu\nu}~\partial_\mu\phi^A~\partial_\nu\phi^B
\end{equation}
where $\gamma_{AB}(\phi)$ is the field metric, $\eta^{\mu\nu}=\text{diag}(1,-1,-1,-1)$ is the Minkowski tensor, $\lambda$ is a scaling constant with dimensions of (length/energy)$^{1/2}$ and $\phi={\phi^A}$ is the collection of fields. Greek indices run from 0 to $d-1$, where $d$ is the dimension of the space-time, and upper-case Latin indices run from 1 to $N$.  

The simplest example of a nonlinear sigma model is the $O(N)$ model, which consists of $N$ real scalar fields, $\phi^A$, $\phi^B$, with the Lagrangian density \cite{hans02}
\begin{equation}\label{2}
\mathcal{L}=\frac{1}{2\lambda^2}~\delta_{AB}~\eta^{\mu\nu}~\frac{\partial\phi^A}{\partial x^\mu}~\frac{\partial\phi^B}{\partial x^\nu}
\end{equation}
where the scalar fields, $\phi^A$, $\phi^B$, satisfy the constraint
\begin{equation}\label{3}
\delta_{AB}~\phi^A\phi^B=1
\end{equation}
and $\delta_{AB}$ is the Kronecker delta. 

The Lagrangian density (\ref{2}) is obviously invariant under the global (space independent) orthogonal transformations $O(N)$, i.e. the group of $N$-dimensional rotations \cite{hans02}
\begin{equation}\label{4}
\phi^A\rightarrow\phi'^A=O^A_B~\phi^B.
\end{equation}
One of the most interesting examples of a $O(N)$ nonlinear sigma model, due to its topological properties, is the $O(3)$ nonlinear sigma model in 1+1 dimensions, with the Lagrangian density 
\begin{equation}\label{5}
\mathcal{L}=\frac{1}{2\lambda^2}~\eta^{\mu\nu}~\partial_\mu\phi~.~\partial_\nu\phi 
\end{equation}
where $\mu$ and $\nu$ range over $\{0,1\}$, and $\phi=(\phi^1,\phi^2,\phi^3)$, subject to the constraint $\phi\cdot\phi=1$, where the dot (.) denotes the standard inner product on real coordinate space of three dimensions, $R^3$. For a $O(3)$ nonlinear sigma model in any number $d$ of space-time dimensions, the target manifold is the unit sphere $S^2$ in $R^3$, and $\mu$ and $\nu$ in the Lagrangian density (\ref{5}) run from 0 to $d-1$.

A simple representation of $\phi$ (in the general time-dependent case) is
\begin{equation}\label{6}
\phi=
\begin{pmatrix}
\sin f(t,{\bf r})~\sin g(t,{\bf r}) \\
\sin f(t,{\bf r})~\cos g(t,{\bf r}) \\
\cos f(t,{\bf r})
\end{pmatrix}
\end{equation}
where $f$ and $g$ are scalar functions on the background space-time, with Minkowski coordinates $x^\mu=(t,{\bf r})$. In what follows, the space-time dimension, $d$, is taken to be 4, and so $\bf r$ is a 3-vector.

If we substitute (\ref{6}) into the Lagrangian density (\ref{5}), then it becomes 
\begin{equation}\label{7}
\mathcal{L}=\frac{1}{2\lambda^2}[\eta^{\mu\nu}~\partial_\mu f~\partial_\nu f+(\sin^2f)~\eta^{\mu\nu}~\partial_\mu g~\partial_\nu g]
\end{equation}
The Euler-Lagrange equations associated with $\mathcal{L}$ in (\ref{7}) are 
\begin{eqnarray}\label{8}
\eta^{\mu\nu}~\partial_\mu\partial_\nu f-(\sin f~\cos f)~\eta^{\mu\nu}~\partial_\mu g~\partial_\nu g=0
\end{eqnarray}
and
\begin{eqnarray}\label{9}
\eta^{\mu\nu}~\partial_\mu\partial_\nu g+2(\cot f)~\eta^{\mu\nu}~\partial_\mu f~\partial_\nu g=0.
\end{eqnarray}

\section{Soliton Solution}
Two solutions to the $O(3)$ field equations (\ref{8}) and (\ref{9}) are 
\begin{itemize}
\item[(i)] a monopole solution, which has form
\begin{equation}\label{10}
\phi=\hat{\textbf{r}}=
\begin{pmatrix}
x/\rho\\
y/\rho\\
z/\rho\\
\end{pmatrix}
\end{equation}
where $\rho=(x^2+y^2+z^2)^{1/2}$ is the spherical radius; and
\item[(ii)] a vortex solution, which is found by imposing the 2-dimensional ''hedgehog'' ansatz
\begin{equation}\label{11}
\phi=
\begin{pmatrix}
\sin f(r)~\sin (n\theta-\chi)\\
\sin f(r)~\cos (n\theta-\chi)\\
\cos f(r)
\end{pmatrix}
\end{equation}
where $r=(x^2+y^2)^{1/2}$, $\theta=\arctan (x/y)$, $n$ is a positive integer, and $\chi$ is a constant phase factor. In this thesis, we only consider the vortex solution.
\end{itemize}

A vortex is a stable time-independent solution to a set of classical field equations that has finite energy in two spatial dimensions; it is a two-dimensional soliton. In three spatial dimensions, a vortex becomes a string, a classical solution with finite energy per unit length \cite{preskill}. Solutions with finite energy, satisfying the appropriate boundary conditions, are candidate soliton solutions \cite{manton}. 

The boundary conditions that are normally imposed on the vortex solution (\ref{11}) are $f(0)=\pi$ and $\lim_{r\to\infty}f(r)=0$, so that the vortex ''unwinds'' from $\phi=-\hat{\textbf{z}}$ to $\phi=\hat{\textbf{z}}$ as $r$ increases from 0 to $\infty$. The function $f$ in this case satisfies the field equation 
\begin{equation}\label{12}
r~\frac{d^2f}{dr^2}+\frac{df}{dr}-\frac{n^2}{r}~\sin f~\cos f=0
\end{equation}
There is in fact a family of solutions to this equation (\ref{12}) satisfying the standard boundary conditions 
\begin{equation}\label{13}
\sin f=\frac{2K^{1/2}r^n}{Kr^{2n}+1}
\end{equation}
or equivalently
\begin{equation}\label{14}
\cos f=\frac{Kr^{2n}-1}{Kr^{2n}+1}
\end{equation}
where $K$ is positive constant.

The energy density, $\sigma$, of a static (time-independent) field with Lagrangian density, $\mathcal{L}$, (\ref{7}) is
\begin{eqnarray}\label{15}
\sigma 
&=& -\mathcal{L} \nonumber\\
&=& \frac{1}{2\lambda^2}\left[\eta^{\mu\nu}~\partial_\mu f~\partial_\nu f+(\sin^2 f)~\eta^{\mu\nu}~\partial_\mu g~\partial_\nu g\right]
\end{eqnarray}
The energy density of the vortex solution is
\begin{eqnarray}\label{17}
\sigma =\frac{4Kn^2}{\lambda^2}\frac{r^{2n-2}}{(Kr^{2n}+1)^2}
\end{eqnarray}
The total energy
\begin{equation}\label{18}
E=\int\int\int \sigma~ dx~dy~dz,
\end{equation}
of the vortex solution is infinite. But, the energy per unit length of the vortex solution
\begin{eqnarray}\label{19}
\mu
&=& \int\int \sigma~dx~dy=2\pi\int_0^\infty\frac{4Kn^2}{\lambda^2}\frac{r^{2n-2}}{(Kr^{2n}+1)^2}~r~dr  \nonumber\\
&=& \frac{4\pi n}{\lambda^2}
\end{eqnarray}
is finite, and does not depend on the value of $K$. (We use the same symbol for the energy per unit length and the mass per unit length, due to the equivalence of energy and mass embodied in the relation $E=mc^2$. Here, we choose units in which $c=1$). 

This last fact means that the vortex solutions in the nonlinear sigma models have no preferred scale. A small value of $K$ corresponds to a more extended vortex solution, and a larger value of $K$ corresponds to a more compact vortex solution, as can be seen by plotting $f$ (or $-\mathcal{L}$) for different values of $K$ and a fixed value of $n$. This means that the vortex solutions are what is called neutrally stable to changes in scale. As $K$ changes, the scale of the vortex changes, but the mass per unit length, $\mu$, does not. Note that because of equation (\ref{19}), there is a preferred winding number, $n=1$, corresponding to the smallest possible positive value of $\mu$.

Furthermore, it can be shown that the topological charge, $T$, of the vortex defined by
\begin{eqnarray}\label{20}
T\equiv \frac{1}{4\pi}~\varepsilon_{ABC}\int\int \phi^A~\partial_x \phi^B ~\partial_y \phi^C~dx~dy
\end{eqnarray}
where $\varepsilon_{ABC}$ is the Levi-Civita symbol, is conserved, in the sense that $\partial_t T=0$ no matter what coordinate dependence is assumed for $f$ and $g$ in (\ref{11}). 

So, the topological charge is a constant, even when the vortex solutions are perturbed. Also, it is simply shown that for the vortex solutions 
\begin{eqnarray}\label{21}
T
&=&-\frac{1}{\pi}n~[f(\infty)-f(0)]= -\frac{1}{\pi}n~(0-\pi) = n
\end{eqnarray}
and so, the winding number is just the topological charge. Because, there is no natural size for the vortex solutions, we can attempt to stabilize them by adding a Skyrme term to the Lagrangian density.  For compact twisting solutions such as the twisted baby Skyrmion string \cite{nitta1}, in addition to the topological charge, $n$, there is a second conserved quantity called the Hopf charge \cite{nitta1}, \cite{mif55}. 

\section{Skyrmion Vortex without a Twist} 
The original sigma model Lagrangian density (with the unit sphere as target manifold) is
\begin{eqnarray}\label{22}
\mathcal{L}_1=\frac{1}{2\lambda^2}~\eta^{\mu\nu}~\partial_\mu\phi~.~\partial_\nu\phi
\end{eqnarray}
If a Skyrme term is added to (\ref{22}), the result is a modified Lagrangian density
\begin{eqnarray}\label{23}
\mathcal{L}_2
&=&\frac{1}{2\lambda^2}~\eta^{\mu\nu}~\partial_\mu\phi~.~\partial_\nu\phi \nonumber\\
&&-~K_s~\eta^{\kappa\lambda}~\eta^{\mu\nu}(\partial_\kappa\phi\times\partial_\mu\phi)~.~(\partial_\lambda\phi\times\partial_\nu\phi) 
\end{eqnarray}
where the Skyrme term is the second term on the right hand side of (\ref{23}). Here, $K_s$ is a positive coupling constant.

With the choice of field representation (\ref{6}), equation (\ref{23}) becomes  
\begin{eqnarray}\label{24}
\mathcal{L}_2
&=&\frac{1}{2\lambda^2}\left(\eta^{\mu\nu}~\partial_\mu f~\partial_\nu f+\sin^2f~\eta^{\mu\nu}~\partial_\mu g~\partial_\nu g\right) \nonumber\\
&&-~K_s\left[2\sin^2f\left(\eta^{\mu\nu}~\partial_\mu f~\partial_\nu f\right)\left(\eta^{\kappa\lambda}~\partial_\kappa g~\partial_\lambda g\right) \right.\nonumber\\
&&\left.-~2\sin^2f\left(\eta^{\mu\nu}~\partial_\mu f~\partial_\nu g\right)^2\right]
\end{eqnarray}
If the vortex configuration (\ref{11}) for $\phi$ is assumed, the Lagrangian density (\ref{7}) becomes 
\begin{eqnarray}\label{25}
\mathcal{L}
&=& -\frac{1}{2\lambda^2}\left[\left(\frac{df}{dr}\right)^2 +\frac{n^2}{r^2}\sin^2f\right] \nonumber\\
&&-~2K_s\frac{n^2}{r^2}\sin^2f\left(\frac{df}{dr}\right)^2
\end{eqnarray}

The Euler-Lagrange equations generated by $\mathcal{L}_2$ (\ref{24}), namely
\begin{eqnarray}\label{26}
\partial_\alpha\left[\frac{\partial\mathcal{L}_2}{\partial(\partial_\alpha f)}\right]  -\frac{\partial\mathcal{L}_2}{\partial f}=0
\end{eqnarray}
and
\begin{eqnarray}\label{27}
\partial_\alpha\left[\frac{\partial\mathcal{L}_2}{\partial(\partial_\alpha g)}\right]  -\frac{\partial\mathcal{L}_2}{\partial g}=0
\end{eqnarray}
Reduce to a single second-order equation for $f$ 
\begin{eqnarray}\label{28}
0
&=& \frac{1}{\lambda^2}\left(\frac{d^2f}{dr^2}+\frac{1}{r}\frac{df}{dr}-\frac{n^2}{r^2}\sin f\cos f\right) \nonumber\\
&&+~4K_s~\frac{n^2}{r^2}~\sin^2f\left(\frac{d^2f}{dr^2}-\frac{1}{r}\frac{df}{dr}\right) \nonumber\\
&&+~4K_s~\frac{n^2}{r^2}~\sin f~\cos f\left(\frac{df}{dr}\right)^2
\end{eqnarray}
with the boundary conditions $f(0)=\pi$ and $\lim_{r\rightarrow\infty}f(r)=0$ as before.

If a suitable vortex solution $f(r)$ of this equation exists, it should have a series expansion for $r<<1$ of the form
\begin{eqnarray}\label{29}
f = \pi +ar+br^3+...~~\text{if}~n=1
\end{eqnarray}
or
\begin{eqnarray}\label{30}
f = \pi +ar^n+br^{3n-2}+...~~\text{if}~n\geq 2
\end{eqnarray}
where $a<0$ and $b$ are constants, and for $r>>1$ the asymptotic form
\begin{eqnarray}\label{31}
f = Ar^{-n}-\frac{1}{12}A^3r^{-3n} + ...
\end{eqnarray}
for some constant $A>0$. 

However, it turns out that it is not possible to match these small-distance and large-distance expansions if $K_s\neq 0$: meaning that any solution $f$ of (\ref{28}) either diverges at $r=0$ or as $r\rightarrow\infty$. This result follows from the following simple scaling argument.

Suppose that $f(r)$ is a solution of equation (\ref{28}). Let $q$ be any positive constant and define $f_q(r)\equiv f(qr)$. Substituting $f_q$ in place of $f$ in equation (\ref{28}) gives a value of $\mu$ which depends in general on the value of $q$
\begin{eqnarray}\label{32}
\mu_q
&=&\int\int\left\{\frac{1}{2\lambda^2}\left[\left(\frac{df_q}{dr}\right)^2+\frac{n^2}{r^2}\sin^2f_q\right] \right.\nonumber\\
&&\left.-2K_s\frac{n^2}{r^2}\sin^2f_q\left(\frac{df_q}{dr}\right)^2\right\}r~dr~d\theta
\end{eqnarray}
where
\begin{eqnarray}\label{33}
\frac{df_q}{dr} =qf'(qr)
\end{eqnarray}

So, if $r$ is replaced as the variable of integration by $\overline{r}=qr$, we have
\begin{eqnarray}\label{34}
\mu_q
&=&\int\int\left\{\frac{1}{2\lambda^2}\left[\left(\frac{df(\overline{r})}{d\overline{r}}\right)^2+\frac{n^2}{\overline{r}^2}\sin^2f(\overline{r})\right] \right.\nonumber\\
&&\left.+~2q^2K_s\frac{n^2}{\overline{r}^2}\sin^2f(\overline{r})\left(\frac{df(\overline{r})}{d\overline{r}}\right)^2\right\}\overline{r}~d\overline{r}~d\theta
\end{eqnarray}
In particular,
\begin{eqnarray}\label{35}
\left.\frac{\partial\mu_q}{\partial q}\right|_{q=1}
&=& 4qK_s\int\int\frac{n^2}{\overline{r}^2}\sin^2f(\overline{r})\left(\frac{df(\overline{r})}{d\overline{r}}\right)^2\overline{r}d\overline{r}d\theta\nonumber\\
&>& 0
\end{eqnarray}
But, if $f$ is a localized solution of eq.(\ref{28}), meaning that it remains suitably bounded as $r\rightarrow 0$ and as $r\rightarrow\infty$, it should be a stationary point of $\mu$, meaning that $\partial\mu_q/\partial q|_{q=1}=0$. 

It follows therefore that no localized solution of (\ref{28}) exists. A more rigorous statement of this property follows on from Derrick's theorem \cite{derrick}, which states that a necessary condition for vortex stability is that 
\begin{eqnarray}\label{36}
\left.\frac{\partial \mu}{\partial q}\right|_{q=1}&=&0
\end{eqnarray}
It is evident that (\ref{35}) does not satisfy this criterion.

In an attempt to fix this problem, we could add a ''mass'' term i.e. $K_v(1-\hat{\textbf{z}}.\phi)$, to the Lagrangian density, $\mathcal{L}_2$, where $\hat{\textbf{z}}$ is the direction of $\phi$ at $r=\infty$ (where $f(r)=0$). The Lagrangian density then becomes
\begin{eqnarray}\label{37}
\mathcal{L}_3=\mathcal{L}_2+K_v(1-\underline{n}.\hat{\underline{\phi}})
\end{eqnarray}
[This Lagrangian density corresponds to the baby Skyrmion model in equation (2.2), p.207 of \cite{piette}]. 

The kinetic term (in the case of a free particle) together with the Skyrme term in $\mathcal{L}_2$ are not sufficient to stabilize a baby Skyrmion, as the kinetic term in $2+1$ dimensions is conformally (scale) invariant and the baby Skyrmion can always reduce its energy by inflating indefinitely. This is in contrast to the usual Skyrme model, in which the Skyrme term prohibits the collapse of the $3+1$ soliton \cite{gisiger}. The mass term is added to limit the size of the baby Skyrmion. 

\section{Skyrmion Vortex with a Twist}
Instead of adding a mass term to stabilize the vortex, we will retain the baby Skyrme model Lagrangian (\ref{24}) but include a twist in the field, $g$, in (\ref{11}). That is, instead of choosing \cite{simanek}, \cite{cho}
\begin{equation}\label{38}
g=n\theta-\chi
\end{equation}
we choose
\begin{equation}\label{39}
g=n\theta+mkz
\end{equation}
where $mkz$ is the twist term, $m$ and $n$ are integers, $2\pi/k$ is the period in the $z$-direction.

The Lagrangian density (\ref{24}) then becomes 
\begin{eqnarray}\label{40}
\mathcal{L}_2
&=& \frac{1}{2\lambda^2}\left[\left(\frac{df}{dr}\right)^2+\sin^2f\left(\frac{n^2}{r^2}+m^2k^2\right)\right] \nonumber\\
&&+~2K_s\sin^2f\left(\frac{df}{dr}\right)^2\left(\frac{n^2}{r^2}+m^2k^2\right)
\end{eqnarray}
The value of the twist lies in the fact that in the far field, where $r\to\infty$ then $f\to0$, the Euler-Lagrange equations for $f$ for both $\mathcal{L}_3$ (without a twist) and $\mathcal{L}_2$ (with a twist) are formally identical to leading order, with $m^2k^2/\lambda^2$ in the twisted case playing the role of the mass coupling constant, $K_v$. So, it is expected that the twist term will act to stabilize the vortex just as the mass term does in $\mathcal{L}_3$.

On a physical level, the twist can be identified with a circular stress in the plane, perpendicular to the vortex string (which can be imagined e.g. as a rod aligned with the $z$-axis). The direction of the twist can be clockwise or counter-clockwise. In view of the energy-mass relation, the energy embodied in the stress term contributes to the gravitational field of the string, with the net result that the trajectories of freely-moving test particles differ according to whether they are directed clockwise or counter-clockwise around the string.

The Euler-Lagrange equation corresponding to the twisted Skyrmion string Lagrangian density (\ref{40}) reads
\begin{eqnarray}\label{41}
0
&=&\frac{1}{\lambda^2}\left[\frac{d^2f}{dr^2}+\frac{1}{r}~\frac{df}{dr}-\left(\frac{n^2}{r^2}+m^2k^2\right)~\sin f~\cos f\right] \nonumber\\
&&+~4K_s~\frac{n^2}{r^2}~\sin^2f\left(\frac{d^2f}{dr^2} -\frac{1}{r}\frac{df}{dr}\right) \nonumber\\
&&+~4K_s~m^2k^2\sin^2f\left(\frac{d^2f}{dr^2}+\frac{1}{r}\frac{df}{dr}\right)\nonumber\\
&&+~4\left(\frac{n^2}{r^2}+m^2k^2\right)K_s~\sin f~\cos f\left(\frac{df}{dr}\right)^2
\end{eqnarray}
It should be noted that the second Euler-Lagrange equation (\ref{27}) is satisfied identically if $g$ has the functional form (\ref{39}).

\section{The Gravitational Field of a Twisted Skyrmion String}
We are interested in constructing the space-time generated by a twisted Skyrmion string. Without gravity, the Lagrangian density of the system is $\mathcal{L}_2$, as given in equation (\ref{24}).
To add gravity, we replace $\eta^{\mu\nu}$ in $\mathcal{L}_2$ with a space-time metric tensor, $g^{\mu\nu}$, which in view of the time-independence and cylindrical symmetry of the assumed vortex solution is taken to be a function of $r$ alone. 

The contravariant metric tensor, $g^{\mu\nu}$, is of course the inverse of the covariant metric tensor, $g_{\mu\nu}$, of the space-time meaning that $g^{\mu\nu}=(g_{\mu\nu})^{-1}$. We use a cylindrical coordinate system $(t,r,\theta,z)$, where $t$ and $z$ have unbounded range, $r\in[0,\infty)$ and $\theta\in[0,2\pi)$. 

The components of the metric tensor
\begin{eqnarray}\label{74}
g_{\mu\nu}
=
\begin{pmatrix}
g_{tt} & 0      & 0                & 0  \\
0      & g_{rr} & 0                & 0  \\
0      & 0      & g_{\theta\theta} & g_{\theta z} \\
0      & 0     & g_{z\theta}      & g_{zz} \\
\end{pmatrix}
\end{eqnarray}
are all functions of $r$, and the presence of the off-diagonal components $g_{\theta z}=g_{z\theta}$ reflects the twist in the space-time (see the twist term, $mkz$, in previous equation i.e. $g=n\theta +mkz$).

The Lagrangian we will be using is
\begin{eqnarray}\label{75}
\mathcal{L}_4
&=&\frac{1}{2\lambda^2}(g^{\mu\nu}~\partial_\mu f~\partial_\nu f+\sin^2f~ g^{\mu\nu}~\partial_\mu g~\partial_\nu g) \nonumber\\
&&-~2K_s\sin^2f~[(g^{\mu\nu}~\partial_\mu f~\partial_\nu f)(g^{\kappa\lambda}~\partial_\kappa g~\partial_\lambda g) \nonumber\\
&&-~2\sin^2f~(g^{\mu\nu}~\partial_\mu f~\partial_\nu g)^2]
\end{eqnarray}
where $f=f(r)$ and $g=n\theta+mkz$.

We need to solve:
\begin{itemize}
\item[(i)] the Einstein equations
\begin{eqnarray}\label{76}
G_{\mu\nu}
&=& -\frac{8\pi G}{c^4}~T_{\mu\nu}  
\end{eqnarray}
where the stress-energy tensor of the vortex, $T_{\mu\nu}$, is defined by
\begin{eqnarray}\label{77}
T_{\mu\nu}
&\equiv&2\frac{\partial\mathcal{L}_4}{\partial g^{\mu\nu}}-g_{\mu\nu}~\mathcal{L}_4
\end{eqnarray}
and
\begin{eqnarray}\label{78}
G_{\mu\nu} 
&=& R_{\mu\nu}-\frac{1}{2}g_{\mu\nu}~R 
\end{eqnarray}
with
$R^{\mu\nu}$ the Ricci tensor and 
\begin{eqnarray}\label{79}
R=g_{\mu\nu}~R^{\mu\nu} = g^{\mu\nu}~R_{\mu\nu}
\end{eqnarray}
the Ricci scalar; and
\item[(ii)] the field equations for $f$ and $g$
\begin{eqnarray}\label{80}
\nabla^\mu\frac{\partial\mathcal{L}_4}{\partial(\partial f/\partial x^\mu)}=\frac{\partial\mathcal{L}_4}{\partial f};~~\nabla^\mu\frac{\partial\mathcal{L}_4}{\partial(\partial g/\partial x^\mu)}=\frac{\partial\mathcal{L}_4}{\partial g}
\end{eqnarray}
\end{itemize}

However, the field equations for $f$ and $g$ are in fact redundant, as they are satisfied identically whenever the Einstein equations are satisfied, by virtue of the Bianchi identities (i.e. permuting of the covariant derivative of the Riemann tensor) $\nabla_\mu G^{\mu}_{\nu}=0$. So, only the Einstein equations will be considered in this section.

To simplify the Einstein equations, we first choose a gauge condition that narrows down the form of the metric tensor. The gauge condition preferred here is that
\begin{eqnarray}\label{81}
g_{\theta\theta}~g_{zz}-(g_{\theta z})^2=r^2
\end{eqnarray}
The geometric significance of this choice is that the determinant of the 2-metric tensor projected onto the surfaces of constant $t$ and $z$ is $r^2$, and so the area element on these surfaces is just $r~dr~d\theta$.

As a further simplification, we write
\begin{eqnarray}\label{82}
g_{tt}=A^2;~~g_{rr}=-B^2;~~g_{\theta\theta}=-C^2;~~g_{\theta z}=\omega
\end{eqnarray}
where $A(r)$, $B(r)$, $C(r)$, $\omega(r)$ and so
\begin{eqnarray}\label{83}
g_{zz}=-\left(\frac{r^2+\omega^2}{C^2}\right).
\end{eqnarray}
The metric tensor, $g_{\mu\nu}$, therefore has the form
\begin{eqnarray}\label{84}
g_{\mu\nu}=
\begin{pmatrix}
A^2  &  0     &    0     &  0  \\
0    &  -B^2  &    0     &  0  \\
0    &  0     &  -C^2    &  \omega  \\
0    &  0     &   \omega & -\left(\frac{r^2+\omega^2}{C^2}\right)
\end{pmatrix}
\end{eqnarray}

\section{Einstein Field Equations}
The Einstein tensor, $G_{\mu\nu}$, is defined as
\begin{equation}\label{210}
G_{\mu\nu} \equiv R_{\mu\nu} - \frac{1}{2}g_{\mu\nu}~R
\end{equation}
where $R_{\mu\nu}$ is the Ricci curvature tensor, $R$ is the Ricci scalar and $g_{\mu\nu}$ is the metric tensor. 

Using (\ref{79}) then (\ref{210}) can be rewritten as 
\begin{eqnarray}\label{210.1}
G_{\mu\nu}\nonumber
&=& R_{\mu\nu} - \frac{1}{2}g_{\mu\nu}~R = R_{\mu\nu} - \frac{1}{2}g_{\mu\nu}~g^{\alpha\beta}~R_{\alpha\beta}\nonumber\\
&=& \delta^{\alpha}_{\mu}~\delta^{\beta}_{\nu}~R_{\alpha\beta} - \frac{1}{2}g_{\mu\nu}~g^{\alpha\beta}~R_{\alpha\beta} \nonumber\\
&=& \big(\delta^{\alpha}_{\mu}~\delta^{\beta}_{\nu} - \frac{1}{2}g_{\mu\nu}~g^{\alpha\beta}\big)R_{\alpha\beta}\nonumber\\
&=& \big(\delta^{\alpha}_{\mu}~\delta^{\beta}_{\nu} - \frac{1}{2}g_{\mu\nu}~g^{\alpha\beta}\big)\nonumber\\
&&\times\big(\Gamma^{\rho}_{\alpha\beta,\rho}-\Gamma^{\rho}_{\alpha\rho,\beta}+\Gamma^{\rho}_{\rho\lambda}\Gamma^{\lambda}_{\alpha\beta}-\Gamma^{\rho}_{\beta\lambda}\Gamma^{\lambda}_{\rho\alpha}\big)
\end{eqnarray}
where $\delta^{\alpha}_{\mu}$, $\delta^{\beta}_{\nu}$ are Kronecker deltas,
\begin{eqnarray}\label{210.2}
\delta^{\alpha}_{\mu}
=\Big\{
\begin{matrix}
1~~~~~\text{if $\alpha = \mu$,} \\
0~~~~~\text{if $\alpha \not= \mu$,} \\
\end{matrix}~~~~~~~~~~
\delta^{\beta}_{\nu}
=\Big\{
\begin{matrix}
1~~~~~\text{if $\beta = \nu$} \\
0~~~~~\text{if $\beta \not= \nu$} \\
\end{matrix}
\end{eqnarray}

We now calculate (\ref{210}) using covariant metric tensor below
\begin{eqnarray}\label{210.4}
g_{tt}
&=&A(r)^2;~~~g_{rr}=-B(r)^2;~~~g_{\theta\theta} = -C(r)^2 \nonumber\\
g_{\theta z}&=&g_{z\theta}=\omega (r);~~~g_{zz}=-\left(\frac{r^2+\omega^2}{C^2}\right)
\end{eqnarray}
From (\ref{210.1}) we obtain 
\begin{eqnarray}\label{210.5}
G_{tt}
&=& \frac{A^2C'^2}{B^2C^2}\Big(1+\frac{\omega^2}{r^2}\Big) - \frac{A^2B'}{rB^3} - \frac{\omega A^2\omega'C'}{r^2B^2C} \nonumber\\
&&-~\frac{A^2C'}{rB^2C} + \frac{A^2\omega'^2}{4r^2B^2} 
\end{eqnarray}
\begin{eqnarray}\label{211}
G_{rr}
&=& \frac{C'^2}{C^2}\Big(1+\frac{\omega^2}{r^2}\Big) -\frac{\omega\omega'C'}{r^2C} -\frac{C'}{rC} -\frac{A'}{rA} 
+\frac{\omega'^2}{4r^2}
\end{eqnarray}
\begin{eqnarray}\label{212}
G_{\theta\theta}
&=& R_{\theta\theta} -\frac{1}{2}g_{\theta\theta}~R = R_{\theta\theta} -\frac{1}{2}(-C^2)~R \nonumber\\
&=& R_{\theta\theta} + \frac{C^2R}{2}
\end{eqnarray}
Substituting $R_{\theta\theta}$ and $R$ into (\ref{212}), we obtain
\begin{eqnarray}\label{213}
G_{\theta\theta}
&=& \frac{C^2}{B^2}\Big[\frac{C''}{C} - \frac{A''}{A} + \frac{A'C'}{AC} - \frac{B'C'}{BC} + \frac{A'B'}{AB}  - \frac{A'}{rA} \nonumber\\
&& +\frac{C'^2}{C^2} - \frac{2B'}{rB} - \frac{C'}{rC}\Big] - \frac{3C^2}{B^2}\Big[\frac{C'^2}{C^2}\Big(1+\frac{\omega^2}{r^2}\Big) \nonumber\\
&& - \frac{\omega \omega' C'}{r^2C} - \frac{C'}{rC} - \frac{B'}{rB} + \frac{\omega'^2}{4r^2}\Big]
\end{eqnarray}
\begin{eqnarray}\label{214}
G_{\theta z}
&=& R_{\theta z} - \frac{\omega R}{2}
\end{eqnarray}
Substituting $R_{\theta z}$ and $R$ into (\ref{214}), we obtain 
\begin{eqnarray}\label{215}
G_{\theta z}
&=& \frac{\omega}{B^2}\Big[\frac{A''}{A} - \frac{\omega''}{2\omega} - \frac{A'\omega'}{2\omega A} + \frac{B'\omega'}{2\omega B} - \frac{A'B'}{AB} + \frac{A'}{rA} \nonumber\\
&& +\frac{\omega'}{2\omega r} +\frac{2B'}{rB}\Big] +\frac{3\omega}{B^2}\Big[ \frac{C'^2}{C^2}\Big(1+\frac{\omega^2}{r^2}\Big) - \frac{\omega\omega'C'}{r^2C} \nonumber\\
&& - \frac{C'}{rC} - \frac{B'}{rB} + \frac{\omega'^2}{4r^2} \Big]
\end{eqnarray}
\begin{eqnarray}\label{216}
G_{zz}
&=& -\frac{\omega^2A'}{rAB^2C^2} + \frac{\omega^2 B'}{rB^3C^2} + \frac{4\omega^2C'}{rB^2C^3} - \frac{\omega\omega'}{rB^2C^2} \nonumber\\
&& + \frac{\omega\omega' A'}{AB^2C^2} - \frac{\omega\omega' B'}{B^3C^2}  +\frac{\omega\omega'C'}{B^2C^3}\Big(\frac{3\omega^2}{r^2}-1\Big)  \nonumber\\
&& + \frac{\omega'^2}{4B^2C^2}\Big(1-\frac{3\omega^2}{r^2}\Big) + \frac{r^2A'B'}{AB^3C^2}\Big(1+\frac{\omega^2}{r^2}\Big)\nonumber\\
&& -\frac{r^2A'C'}{AB^2C^3}\Big(1+\frac{\omega^2}{r^2}\Big) + \frac{r^2B'C'}{B^3C^3}\Big(1+\frac{\omega^2}{r^2}\Big) \nonumber\\
&& - \frac{r^2A''}{AB^2C^2}\Big(1+\frac{\omega^2}{r^2}\Big)  - \frac{r^2C''}{B^2C^3}\Big(1+\frac{\omega^2}{r^2}\Big) \nonumber\\
&& + \frac{\omega\omega''}{B^2C^2} - \frac{3\omega^2C'^2}{B^2C^4}\Big(1+\frac{\omega^2}{r^2}\Big)
\end{eqnarray}
Multiplying (\ref{210.5}) with $B^2/A^2$, we obtain
\begin{eqnarray}\label{217}
\frac{B^2}{A^2}G_{tt}
&=& \frac{C'^2}{C^2}\Big(1+\frac{\omega^2}{r^2}\Big)  - \frac{\omega \omega'C'}{r^2C}-\frac{C'}{rC} \nonumber\\
&& - \frac{B'}{rB} + \frac{\omega'^2}{4r^2}
\end{eqnarray}
Comparing this equation with (\ref{211}), we see that
\begin{eqnarray}\label{218}
G_{rr}
&=& \frac{B'}{rB} -\frac{A'}{rA} +\frac{B^2}{A^2}G_{tt}
\end{eqnarray}
From (\ref{213}) we obtain
\begin{eqnarray}\label{219}
\frac{B^2}{C^2}G_{\theta\theta}
&=& \frac{C''}{C} - \frac{A''}{A} + V' - \frac{3B^2}{A^2}G_{tt}
\end{eqnarray}
where
\begin{eqnarray}\label{220}
V'
&=& \frac{A'C'}{AC} - \frac{B'C'}{BC} + \frac{A'B'}{AB}  - \frac{A'}{rA} +\frac{C'^2}{C^2} \nonumber\\
&& - \frac{2B'}{rB} - \frac{C'}{rC}
\end{eqnarray}
From (\ref{215}) we obtain
\begin{eqnarray}\label{221}
\frac{B^2}{\omega}G_{\theta z}
&=& \frac{A''}{A} - \frac{\omega''}{2\omega} + W' +\frac{3B^2}{A^2}G_{tt}
\end{eqnarray}
where
\begin{eqnarray}\label{222}
W'
&=& - \frac{A'\omega'}{2\omega A} + \frac{B'\omega'}{2\omega B} - \frac{A'B'}{AB} + \frac{A'}{rA} + \frac{\omega'}{2\omega r} \nonumber\\
&& +\frac{2B'}{rB}
\end{eqnarray}
From (\ref{216}) we obtain
\begin{eqnarray}\label{223}
\frac{B^2C^2}{\omega^2}G_{zz}
&=& -\frac{A''}{A}\Big(1+\frac{r^2}{\omega^2}\Big) -\frac{C''}{C}\Big(1+\frac{r^2}{\omega^2}\Big) \nonumber\\
&& +\frac{\omega''}{\omega} +X'  -\frac{3B^2}{A^2}G_{tt}
\end{eqnarray}
where
\begin{eqnarray}\label{224}
X'
&=&  -~\frac{A'}{rA} - \frac{\omega'}{\omega r}+ \frac{\omega'A'}{\omega A} - \frac{\omega'B'}{\omega B} + \frac{A'B'}{AB}\Big(1+\frac{r^2}{\omega^2}\Big) \nonumber\\
&& - \frac{A'C'}{AC}\Big(1+\frac{r^2}{\omega^2}\Big) +\frac{B'C'}{BC}\Big(1+\frac{r^2}{\omega^2}\Big) - \frac{2B'}{rB} \nonumber\\
&& - \frac{\omega'C'}{\omega C} + \frac{C'}{rC} + \frac{\omega'^2}{4\omega^2}
\end{eqnarray}

The Einstein field equations can be defined in the form
\begin{eqnarray}\label{225}
G_{\mu\nu}
&\equiv & -\varepsilon T_{\mu\nu}
\end{eqnarray}
where $\varepsilon = \frac{8\pi G}{c^4}$. 

From (\ref{225}), we obtain
\begin{equation}\label{226}
T_{tt} = \frac{A^2}{2B^2}\Big[\Big(\frac{1}{\lambda^2} + 4NK_s\sin^2f\Big)\left(\frac{\partial f}{\partial r}\right)^2 - \frac{A^2N\sin^2f}{2\lambda^2}\Big]
\end{equation}
\begin{eqnarray}\label{227}
T_{rr} 
&=& \frac{B^2}{A^2}T_{tt} + \frac{B^2N\sin^2f}{\lambda^2} 
\end{eqnarray}
\begin{eqnarray}\label{228}
T_{\theta\theta} 
&=& -\frac{C^2}{A^2}T_{tt} +\left[\frac{1}{\lambda^2} -\frac{4K_s}{B^2}\left(\frac{\partial f}{\partial r}\right)^2\right]n^2\sin^2f
\end{eqnarray}
\begin{eqnarray}\label{229}
T_{\theta z} 
&=& \frac{\omega}{A^2}T_{tt} +\left[\frac{1}{\lambda^2} -\frac{4K_s}{B^2}\left(\frac{\partial f}{\partial r}\right)^2\right]nmk~\sin^2f
\end{eqnarray}
\begin{eqnarray}\label{230}
T_{zz} 
&=& -\frac{r^2+\omega^2}{A^2C^2}T_{tt} +\left[\frac{1}{\lambda^2} -\frac{4K_s}{B^2}\left(\frac{\partial f}{\partial r}\right)^2\right]m^2k^2\sin^2f \nonumber\\
\end{eqnarray}
Eliminating $A''$ from eqs.(\ref{219}) and (\ref{213}) we obtain
\begin{eqnarray}\label{236}
\frac{B^2}{C^2}G_{\theta\theta} + \frac{B^2}{\omega}G_{\theta z} 
&=& \frac{C''}{C} - \frac{\omega''}{2\omega} + V' + W'
\end{eqnarray}
Eliminating $A''$ from eqs.(\ref{213}) and (\ref{223}), we obtain
\begin{eqnarray}\label{237}
&&
\frac{B^2}{\omega}\Big(1+\frac{r^2}{\omega^2}\Big)G_{\theta z} + \frac{B^2C^2}{\omega^2}G_{zz} \nonumber\\
&=& -~\frac{C''}{C}\Big(1+\frac{r^2}{\omega^2}\Big) + \frac{\omega''}{2\omega}\Big(1-\frac{r^2}{\omega^2}\Big)\nonumber\\
&& +~ W'\Big(1+\frac{r^2}{\omega^2}\Big) + X' + \frac{3B^2r^2}{\omega^2A^2}G_{tt}
\end{eqnarray}
Eliminating $C''$ from eqs.(\ref{236}) and (\ref{237}), we obtain 
\begin{eqnarray}\label{238}
\omega'' 
&=& \varepsilon\omega\Big(\frac{B^2}{\lambda^2} - 4K_sf'^2\Big)\Big[\Big(\frac{n^2}{C^2} +  \frac{2nmk}{\omega}\Big)\Big(1+\frac{\omega^2}{r^2}\Big) \nonumber\\
&& +\frac{C^2m^2k^2}{r^2}\Big]\sin^2(f)  +\frac{4\omega C'^2}{C^2}\Big(1+\frac{\omega^2}{r^2}\Big)-\frac{4\omega^2\omega'C'}{r^2C} \nonumber\\
&& -\frac{4\omega C'}{rC} +\frac{\omega\omega'^2}{r^2} +\frac{\omega A'}{rA} - \frac{\omega B'}{rB}  \nonumber\\
&& -\frac{A'\omega'}{A} +\frac{B'\omega'}{B}+\frac{\omega'}{r}
\end{eqnarray}

Substituting eq.(\ref{238}) into (\ref{237}), we obtain 
\begin{eqnarray}\label{239}
C'' 
&=& \frac{\varepsilon C}{2}\Big(\frac{B^2}{\lambda^2}-4K_s f'^2\Big)\Big[\frac{n^2}{C^2}\Big(\frac{\omega^2}{r^2}-1\Big)+\frac{2nmk\omega}{r^2} \nonumber\\
&& + \frac{C^2m^2k^2}{r^2}\Big]\sin^2(f) +\frac{C'^2}{C}\Big(1+\frac{2\omega^2}{r^2}\Big) - \frac{2\omega\omega'C'}{r^2} \nonumber\\
&& -\frac{C'}{r} - \frac{CB'}{2rB} + \frac{C\omega'^2}{2r^2} - \frac{A'C'}{A} + \frac{B'C'}{B} + \frac{CA'}{2rA} \nonumber\\
\end{eqnarray}
Substituting eq.(\ref{239}) into (\ref{236}), we obtain 
\begin{eqnarray}\label{240}
A''
&=& \frac{\varepsilon A}{2}\Big(\frac{B^2}{\lambda^2} - 4K_sf'^2\Big)\Big[\frac{n^2}{C^2}\Big(1+\frac{\omega^2}{r^2}\Big)+\frac{2nmk\omega}{r^2} \nonumber\\
&& +\frac{C^2m^2k^2}{r^2}\Big]\sin^2(f)  +\frac{A'B'}{B} - \frac{A'}{2r} - \frac{AB'}{2rB} 
\end{eqnarray}
Using (\ref{218}), we obtain
\begin{eqnarray}\label{242}
B'
&=& \frac{BA'}{A} + \frac{r\varepsilon B^3}{\lambda^2}\Big[\frac{n^2}{C^2}\Big(1+\frac{\omega^2}{r^2}\Big)  + \frac{2nmk\omega}{r^2} \nonumber\\
&&+\frac{C^2m^2k^2}{r^2}\Big]\sin^2f
\end{eqnarray}
where
\begin{equation}\label{243}
Nr^2 =  -\frac{n^2(r^2+\omega^2)}{C^2} - 2nmk\omega - C^2m^2k^2
\end{equation}
Using (\ref{225}), (\ref{226}) we obtain
\begin{eqnarray}\label{245}
f'
&=&  \Big(\frac{\varepsilon}{2\lambda^2}+2\varepsilon NK_s\sin^2f\Big)^{-1/2}\nonumber\\
&&\times~\Big[\frac{\varepsilon A^2N\sin^2f}{4\lambda^2} -\frac{C'^2}{C^2}\Big(1+\frac{\omega^2}{r^2}\Big) 
\nonumber\\
&&+\frac{\omega\omega'C'}{r^2C} +\frac{C'}{rC} +\frac{B'}{rB} -\frac{\omega'^2}{4r^2}\Big]^{1/2}
\end{eqnarray}
Using (\ref{242}) to eliminate $B'$, eq.(\ref{245}) can be rewritten as
\begin{eqnarray}\label{247}
f'
&=& \left\{\frac{\varepsilon}{2\lambda^2} -2\varepsilon K_s\Big[\frac{n^2}{C^2}\Big(1+\frac{\omega^2}{r^2}\Big) +\frac{2nmk\omega}{r^2} 
\right.\nonumber\\
&&\left. +\frac{C^2m^2k^2}{r^2}\Big]\sin^2f\right\}^{-1/2}\nonumber\\
&& \times~\left\{\frac{A'}{rA}+ \frac{\omega\omega'C'}{r^2C} + \frac{C'}{rC} -\frac{C'^2}{C^2}\Big(1+\frac{\omega^2}{r^2}\Big)  
-\frac{\omega'^2}{4r^2} \right.\nonumber\\
&&\left. +\frac{\varepsilon A^2}{4\lambda^2}
\left[\frac{n^2}{C^2}\left(1+\frac{\omega^2}{r^2}\right) +\frac{2nmk\omega}{r^2} +\frac{C^2m^2k^2}{r^2}\right]\sin^2f\right\}^{1/2} \nonumber\\
\end{eqnarray}
Using (\ref{243}), eq.(\ref{240}) can be rewritten as 
\begin{eqnarray}\label{248}
A'' 
&=& -\frac{A'}{r} +\frac{A'B'}{B} +2\varepsilon K_s AN\sin^2ff'^2
\end{eqnarray}
or
\begin{eqnarray}\label{249}
A'' 
&=& -\frac{A'}{r} + \frac{A'^2}{A} +\varepsilon\sin^2f\Big[\frac{n^2}{C^2}\Big(1+\frac{\omega^2}{r^2}\Big) \nonumber\\
&&+\frac{2nmk\omega}{r^2} + \frac{C^2m^2k^2}{r^2}\Big]\nonumber\\
&& \times \left(\frac{rA'B^2}{\lambda^2} - 2K_s\left\{\frac{A'}{r}+ \frac{\omega A\omega'C'}{r^2C}  \right.\right.\nonumber\\
&&\left.\left.+~\frac{AC'}{rC}  -\frac{AC'^2}{C^2}\left(1+\frac{\omega^2}{r^2}\right)  -\frac{A\omega'^2}{4r^2} \right.\right.\nonumber\\
&&\left.\left. +\frac{\varepsilon AB^2}{2\lambda^2}\Big[\frac{n^2}{C^2}\left(1+\frac{\omega^2}{r^2}\right) +\frac{2nmk\omega}{r^2} 
+\frac{C^2m^2k^2}{r^2}\Big]\right.\right.\nonumber\\
&&\left.\left.\times \sin^2f\right\}\right.\nonumber\\
&&\left.\times \left\{\frac{\varepsilon}{2\lambda^2} - 2\varepsilon K_s\Big[\frac{n^2}{C^2}\Big(1+\frac{\omega^2}{r^2}\Big) 
+\frac{2nmk\omega}{r^2}  \right.\right.\nonumber\\
&&\left.\left.+\frac{C^2m^2k^2}{r^2}\Big]\sin^2f\right\}^{-1}\right)
\end{eqnarray}

In order to separate the equations for the purposes of numerical integration, $B'$ and $f'$ should be eliminated from (\ref{239}). Using (\ref{242}) and (\ref{247}) to eliminate $B'$ and $f'$ gives  
\begin{eqnarray}\label{250}
C'' 
&=& \frac{C'^2}{C}\Big(1+\frac{2\omega^2}{r^2}\Big) -\frac{2\omega\omega'C'}{r^2} -\frac{C'}{r} +\frac{C\omega'^2}{2r^2}\nonumber\\
&& -\frac{\varepsilon B^2\sin^2f}{\lambda^2}\Big[\frac{n^2}{C^2}\Big(1+\frac{\omega^2}{r^2}\Big) +\frac{2nmk\omega}{r^2} 
+\frac{C^2m^2k^2}{r^2}\Big]\nonumber\\
&&\times \Big(\frac{C}{2} -rC'\Big) \nonumber\\
&& +\frac{\varepsilon \sin^2f}{2}\Big[\frac{n^2}{C^2}\Big(\frac{\omega^2}{r^2}-1\Big)+\frac{2nmk\omega}{r^2} +\frac{C^2m^2k^2}{r^2}\Big]\nonumber\\
&&\Big(\frac{CB^2}{\lambda^2}-4K_s \left\{\frac{A'C}{rA} +\frac{\omega\omega'C'}{r^2} +\frac{C'}{r} -\frac{C'^2}{C}\Big(1+\frac{\omega^2}{r^2}\Big) \right.\nonumber\\
&&\left.-\frac{\omega'^2C}{4r^2} +\frac{\varepsilon B^2C\sin^2f}{2\lambda^2}\Big[\frac{n^2}{C^2}\Big(1+\frac{\omega^2}{r^2}\Big) \right.\nonumber\\
&&\left.+\frac{2nmk\omega}{r^2} 
+\frac{C^2m^2k^2}{r^2}\Big]\right\}\left\{\frac{\varepsilon}{2\lambda^2} -2\varepsilon K_s\sin^2f\right.\nonumber\\
&&\left.\times\Big[\frac{n^2}{C^2}\Big(1+\frac{\omega^2}{r^2}\Big) 
+\frac{2nmk\omega}{r^2} +\frac{C^2m^2k^2}{r^2}\Big]\right\}^{-1}\Big)
\end{eqnarray}

Similarly, $B'$ and $f'$ need to be eliminated from (\ref{238}). Using (\ref{242}) and (\ref{247}) to eliminate them gives 
\begin{eqnarray}\label{251}
\omega'' 
&=& \frac{\omega'}{r} + \frac{4\omega C'^2}{C^2}\Big(1+\frac{\omega^2}{r^2}\Big)-\frac{4\omega^2\omega'C'}{r^2C} - \frac{4\omega C'}{rC} +\frac{\omega\omega'^2}{r^2}\nonumber\\
&& - \frac{\varepsilon B^2(\omega - r\omega')\sin^2f}{\lambda^2}\Big[\frac{n^2}{C^2}\Big(1+\frac{\omega^2}{r^2}\Big) + \frac{2nmk\omega}{r^2} \nonumber\\
&& + \frac{C^2m^2k^2}{r^2}\Big] +\varepsilon \sin^2f\Big[\Big(\frac{n^2}{C^2} +\frac{2nmk}{\omega}\Big)\Big(1+\frac{\omega^2}{r^2}\Big) \nonumber\\
&&+\frac{C^2m^2k^2}{r^2}\Big]\Big(\frac{\omega B^2}{\lambda^2} - 4K_s \Big\{\frac{\omega A'}{rA} + \frac{\omega^2\omega'C'}{r^2C}\nonumber\\
&& + \frac{\omega C'}{rC} - \frac{\omega C'^2}{C^2}\Big(1+\frac{\omega^2}{r^2}\Big) - \frac{\omega \omega'^2}{4r^2}\nonumber\\
&& + \frac{\varepsilon \omega B^2\sin^2f}{2\lambda^2}\Big[\frac{n^2}{C^2}\Big(1+\frac{\omega^2}{r^2}\Big) + \frac{2nmk\omega}{r^2} + \frac{C^2m^2k^2}{r^2}\Big]\Big\}\nonumber\\
&&\times \Big\{\frac{\varepsilon}{2\lambda^2} - 2\varepsilon K_s\sin^2f\Big[\frac{n^2}{C^2}\Big(1+\frac{\omega^2}{r^2}\Big) + \frac{2nmk\omega}{r^2} \nonumber\\
&&+\frac{C^2m^2k^2}{r^2}\Big]\Big\}^{-1}\Big)
\end{eqnarray}

The system of equations to be solved is therefore
\begin{eqnarray}\label{252}
B'
&=& \frac{BA'}{A} + \frac{r\varepsilon B^3}{\lambda^2}\Big[\frac{n^2}{C^2}\Big(1+\frac{\omega^2}{r^2}\Big)  + \frac{2nmk\omega}{r^2} \nonumber\\
&&+\frac{C^2m^2k^2}{r^2}\Big]\sin^2f
\end{eqnarray}
\begin{eqnarray}\label{253}
f'
&=& \pm \Big\{\frac{\varepsilon}{2\lambda^2} - 2\varepsilon K_s\Big[\frac{n^2}{C^2}\Big(1+\frac{\omega^2}{r^2}\Big) + \frac{2nmk\omega}{r^2} \nonumber\\
&&+\frac{C^2m^2k^2}{r^2}\Big]\sin^2f\Big\}^{-1/2}\Big\{\frac{A'}{rA}+ \frac{\omega\omega'C'}{r^2C} + \frac{C'}{rC} \nonumber\\
&&-\frac{C'^2}{C^2}\Big(1+\frac{\omega^2}{r^2}\Big)  - \frac{\omega'^2}{4r^2} +\frac{\varepsilon B^2}{2\lambda^2}\Big[\frac{n^2}{C^2}\Big(1+\frac{\omega^2}{r^2}\Big) \nonumber\\
&&+\frac{2nmk\omega}{r^2} + \frac{C^2m^2k^2}{r^2}\Big]\sin^2f\Big\}^{1/2}
\end{eqnarray}
\begin{eqnarray}\label{254}
A'' 
&=& -\frac{A'}{r} + \frac{A'^2}{A} +\varepsilon\sin^2f\Big[\frac{n^2}{C^2}\Big(1+\frac{\omega^2}{r^2}\Big) + 
\frac{2nmk\omega}{r^2} \nonumber\\
&&+\frac{C^2m^2k^2}{r^2}\Big]\Big(\frac{rA'B^2}{\lambda^2} - 2K_s\Big\{\frac{A'}{r}+ \frac{\omega A\omega'C'}{r^2C} + \frac{AC'}{rC} \nonumber\\
&& -\frac{AC'^2}{C^2}\Big(1+\frac{\omega^2}{r^2}\Big)  - \frac{A\omega'^2}{4r^2} + \frac{\varepsilon AB^2}{2\lambda^2}\Big[\frac{n^2}{C^2}\Big(1+\frac{\omega^2}{r^2}\Big) \nonumber\\
&& +\frac{2nmk\omega}{r^2} + \frac{C^2m^2k^2}{r^2}\Big]\sin^2f\Big\} \nonumber\\
&&\times\Big\{\frac{\varepsilon}{2\lambda^2} - 2\varepsilon K_s\Big[\frac{n^2}{C^2}\Big(1+\frac{\omega^2}{r^2}\Big) + \frac{2nmk\omega}{r^2} + \frac{C^2m^2k^2}{r^2}\Big]\nonumber\\
&&\sin^2f\Big\}^{-1}\Big)
\end{eqnarray}
\begin{eqnarray}\label{255}
C'' 
&=& \frac{C'^2}{C}\Big(1+\frac{2\omega^2}{r^2}\Big) - \frac{2\omega\omega'C'}{r^2} - \frac{C'}{r} + \frac{C\omega'^2}{2r^2}\nonumber\\
&& - \frac{\varepsilon B^2\sin^2f}{\lambda^2}\Big[\frac{n^2}{C^2}\Big(1+\frac{\omega^2}{r^2}\Big) + \frac{2nmk\omega}{r^2} + \frac{C^2m^2k^2}{r^2}\Big]\nonumber\\
&&\times \Big(\frac{C}{2} - rC'\Big)  + \frac{\varepsilon \sin^2f}{2}\Big[\frac{n^2}{C^2}\Big(\frac{\omega^2}{r^2}-1\Big)+\frac{2nmk\omega}{r^2} \nonumber\\
&&+\frac{C^2m^2k^2}{r^2}\Big]\Big(\frac{CB^2}{\lambda^2}-4K_s \Big\{\frac{A'C}{rA} + \frac{\omega\omega'C'}{r^2}  +\frac{C'}{r} \nonumber\\
&& -\frac{C'^2}{C}\Big(1+\frac{\omega^2}{r^2}\Big) - \frac{\omega'^2C}{4r^2}  + \frac{\varepsilon B^2C\sin^2f}{2\lambda^2}\Big[\frac{n^2}{C^2}\Big(1+\frac{\omega^2}{r^2}\Big) \nonumber\\
&& +\frac{2nmk\omega}{r^2} + \frac{C^2m^2k^2}{r^2}\Big]\Big\}\Big\{\frac{\varepsilon}{2\lambda^2} - 2\varepsilon K_s \sin^2f\nonumber\\
&&\times \Big[\frac{n^2}{C^2}\Big(1+\frac{\omega^2}{r^2}\Big) + \frac{2nmk\omega}{r^2} + \frac{C^2m^2k^2}{r^2}\Big]\Big\}^{-1}\Big)
\end{eqnarray}
\begin{eqnarray}\label{256}
\omega'' 
&=& \frac{\omega'}{r} + \frac{4\omega C'^2}{C^2}\Big(1+\frac{\omega^2}{r^2}\Big)-\frac{4\omega^2\omega'C'}{r^2C} - \frac{4\omega C'}{rC} +\frac{\omega\omega'^2}{r^2}\nonumber\\
&& - \frac{\varepsilon B^2(\omega - r\omega')\sin^2f}{\lambda^2}\Big[\frac{n^2}{C^2}\Big(1+\frac{\omega^2}{r^2}\Big) + \frac{2nmk\omega}{r^2} \nonumber\\
&& +\frac{C^2m^2k^2}{r^2}\Big]  +\varepsilon \sin^2f \Big[\Big(\frac{n^2}{C^2} +  \frac{2nmk}{\omega}\Big)\Big(1+\frac{\omega^2}{r^2}\Big) \nonumber\\
&&+\frac{C^2m^2k^2}{r^2}\Big] \Big(\frac{\omega B^2}{\lambda^2} - 4K_s \Big\{\frac{\omega A'}{rA} + \frac{\omega^2\omega'C'}{r^2C} +\frac{\omega C'}{rC} \nonumber\\
&& -\frac{\omega C'^2}{C^2}\Big(1+\frac{\omega^2}{r^2}\Big) - \frac{\omega \omega'^2}{4r^2} 
+\frac{\varepsilon \omega B^2\sin^2f}{2\lambda^2}\nonumber\\
&&\times \Big[\frac{n^2}{C^2}\Big(1+\frac{\omega^2}{r^2}\Big) + \frac{2nmk\omega}{r^2} + \frac{C^2m^2k^2}{r^2}\Big]\Big\}\nonumber\\
&&\times \Big\{\frac{\varepsilon}{2\lambda^2} - 2\varepsilon K_s\sin^2f\Big[\frac{n^2}{C^2}\Big(1+\frac{\omega^2}{r^2}\Big) + \frac{2nmk\omega}{r^2} \nonumber\\
&&+\frac{C^2m^2k^2}{r^2}\Big]\Big\}^{-1}\Big)
\end{eqnarray}

The plus/minus sign $\pm$ in eq.(\ref{253}) indicates that there is a choice of signs for $f'$. But since the boundary conditions on $f$ will require that $f(0)=\pi$ and $\lim_{r\rightarrow\infty}f(r)=0$, it is likely that $f'<0$ for all values of $r$. 

\section{The Einstein field equations solutions without Skyrme term}
We start to solve the Einstein field equations by assuming that $K_s=0$ (meaning that there is no Skyrme term in the Lagrangian), then the field equations for $A$, $B$, $C$, $\omega$ and $f$ read
\begin{eqnarray}\label{496}
B'
&=&\frac{BA'}{A} +\frac{r\varepsilon B^3}{\lambda^2}\left[\frac{n^2}{C^2}\left(1+\frac{\omega^2}{r^2} \right) +\frac{2nmk\omega}{r^2} \right.\nonumber\\
&&\left.+~\frac{C^2m^2k^2}{r^2}\right]\sin^2f
\end{eqnarray}
\begin{eqnarray}\label{497}
f'
&=& \pm\left(\frac{\varepsilon}{2\lambda^2}\right)^{-1/2}\left\{\frac{A'}{rA} +\frac{\omega\omega'C'}{r^2C} +\frac{C'}{rC} -\frac{C'^2}{C^2}\left(1+\frac{\omega^2}{r^2}\right) \right.\nonumber\\
&&\left.-~\frac{\omega'^2}{4r^2} +\frac{\varepsilon B^2}{2\lambda^2}\left[\frac{n^2}{C^2}\left(1+\frac{\omega^2}{r^2}\right) +\frac{2nmk\omega}{r^2} +\frac{C^2m^2k^2}{r^2}\right]\right.\nonumber\\
&&\left.\times~ \sin^2f\right\}^{1/2}
\end{eqnarray}
\begin{eqnarray}\label{498}
A''
&=& -\frac{A'}{r} +\frac{A'^2}{A} +\frac{\varepsilon}{\lambda^2}rA'B^2\left[\frac{n^2}{C^2}\left(1+\frac{\omega^2}{r^2}\right) +\frac{2nmk\omega}{r^2} \right.\nonumber\\
&&\left.+~\frac{C^2m^2k^2}{r^2}\right]\sin^2f
\end{eqnarray}
\begin{eqnarray}\label{499}
C''
&=& \frac{C'^2}{C}\left(1+\frac{2\omega^2}{r^2}\right) -\frac{2\omega\omega'C'}{r^2} -\frac{C'}{r} +\frac{C\omega'^2}{2r^2} \nonumber\\
&&-\frac{\varepsilon B^2}{\lambda^2}\left(\frac{C}{2} -rC'\right)\left[\frac{n^2}{C^2}\left(1+\frac{\omega^2}{r^2}\right) +\frac{2nmk\omega}{r^2} \right.\nonumber\\
&&\left.+\frac{C^2m^2k^2}{r^2}\right]\sin^2f +\frac{\varepsilon CB^2}{2\lambda^2}\left[\frac{n^2}{C^2}\left(\frac{\omega^2}{r^2}-1\right) \right.\nonumber\\
&&\left.+\frac{2nmk\omega}{r^2} +\frac{C^2m^2k^2}{r^2}\right]\sin^2f
\end{eqnarray}
and
\begin{eqnarray}\label{500}
\omega''
&=& \frac{\omega'}{r} +\frac{4\omega C'^2}{C^2}\left(1+\frac{\omega^2}{r^2}\right) -\frac{4\omega\omega'C'}{r^2C} -\frac{4\omega C'}{rC} +\frac{\omega\omega'^2}{r^2} \nonumber\\
&& -\frac{\varepsilon B^2}{\lambda^2}(\omega-r\omega')\left[\frac{n^2}{C^2}(\left(1+\frac{\omega^2}{r^2}\right) +\frac{2nmk\omega}{r^2} \right.\nonumber\\
&&\left.+\frac{C^2m^2k^2}{r^2}\right]\sin^2f +\frac{\varepsilon\omega B^2}{\lambda^2}\nonumber\\
&&\times\left[\left(\frac{n^2}{C^2}+\frac{2mnk}{\omega}\right)\left(1+\frac{\omega^2}{r^2}\right) +\frac{C^2m^2k^2}{r^2}\right]\sin^2f \nonumber\\
\end{eqnarray}

A solution that is regular on the axis $r=0$ will have $A=A'=0$ there, and so according to the third equation above $A''=0$ at $r=0$ as well. This means that $A$ will be constant everywhere, and can always be rescaled so that $A\equiv 1$. The first two equations then become
\begin{eqnarray}\label{501}
B'
&=& \frac{r\varepsilon B^3}{\lambda^2}\left[\frac{n^2}{C^2}\left(1 +\frac{\omega^2}{r^2}\right) +\frac{2nmk\omega}{r^2} +\frac{C^2m^2k^2}{r^2}\right]\sin^2f \nonumber\\
\end{eqnarray}
\begin{eqnarray}\label{502}
f'
&=& \pm\left(\frac{\varepsilon}{2\lambda^2}\right)^{-1/2}\left\{\frac{\omega\omega'C'}{r^2C} +\frac{C'}{rC} -\frac{C'^2}{C^2}\left(1 +\frac{\omega^2}{r^2}\right) -\frac{\omega'^2}{4r^2} \right.\nonumber\\
&&\left.+\frac{\varepsilon B^2}{2\lambda^2}\left[\frac{n^2}{C^2}\left(1 +\frac{\omega^2}{r^2}\right) +\frac{2nmk\omega}{r^2} +\frac{C^2m^2k^2}{r^2}\right]\right.\nonumber\\
&&\left.\times\sin^2f\right\}^{1/2}
\end{eqnarray}

\subsection{Non-twisting case ($\omega=0$)}
If we assume further that $mk=0$ and $\omega\equiv 0$ then the three remaining field equations read
\begin{eqnarray}\label{503}
B'
&=& \frac{r\varepsilon B^3}{\lambda^2}\frac{n^2}{C^2}\sin^2f
\end{eqnarray}
\begin{eqnarray}\label{504}
f'
&=& \pm\left(\frac{\varepsilon}{2\lambda^2}\right)^{-1/2}\left(\frac{C'}{rC} -\frac{C'^2}{C^2} +\frac{\varepsilon B^2}{2\lambda^2}\frac{n^2}{C^2}\sin^2f\right)^{1/2}
\end{eqnarray}
and
\begin{eqnarray}\label{505}
C''
&=& -\frac{C'}{rC}(C-rC') -\frac{\varepsilon B^2}{\lambda^2}(C-rC')\frac{n^2}{C^2}\sin^2f
\end{eqnarray}

Again, a solution that is regular on the axis will have $B\approx B_0>0$ and $C\approx B_0r$ for $r$ small. According to the equation for $C$, therefore $C''=0$ at $r=0$ and $C''$ remains 0 as long as $C$ is strictly proportional to $r$. Hence, $C=B_0r$ everywhere, and the field equations for $B$ and $f$ now become
\begin{eqnarray}\label{506}
\left(\frac{B}{B_0}\right)'
&=& \frac{\varepsilon}{\lambda^2}\left(\frac{B}{B_0}\right)^3\frac{n^2}{r}\sin^2f
\end{eqnarray}
and
\begin{eqnarray}\label{507}
f'
&=& \pm\frac{B}{B_0}\frac{n}{r}\sin f
\end{eqnarray}

If the last equation is substituted into the expressions developed for components of the stress-energy tensor $T_{\mu\nu}$, we find that
\begin{eqnarray}\label{508}
T_{tt}
= \frac{A^2}{2\lambda^2}\left(\frac{f'^2}{B^2} +\frac{n^2}{C^2}\sin^2f\right)=\frac{1}{\lambda^2}\frac{n^2}{B_0^2r^2}\sin^2f
\end{eqnarray}
\begin{eqnarray}\label{509}
T_{rr}
&=& \frac{1}{2\lambda^2}\left(f'^2 -\frac{n^2B^2}{C^2}\sin^2f\right)=0
\end{eqnarray}
\begin{eqnarray}\label{510}
T_{\theta\theta}
&=& -\frac{1}{2\lambda^2}\left(\frac{C^2f'^2}{B^2} -n^2\sin^2f\right) = 0
\end{eqnarray}
and
\begin{eqnarray}\label{511}
T_{zz}=-\frac{1}{2\lambda^2}\left(\frac{r^2f'^2}{B^2C^2} +\frac{n^2r^2}{C^4}\sin^2f\right) = -\frac{1}{\lambda^2}\frac{n^2}{B_0^4r^2}\sin^2f \nonumber\\
\end{eqnarray}
Also, the line element in this case is
\begin{eqnarray}\label{512}
ds^2
&=& dt^2 -B(r)^2dr^2 -B_0^2r^2d\theta^2 -B_0^{-2}dz^2
\end{eqnarray}
so
\begin{eqnarray}\label{513}
T_t^t
&=& \frac{1}{\lambda^2}\frac{n^2}{B_0^2r^2}\sin^2f = T_z^z
\end{eqnarray}

Cylindrical space-times with a stress-energy content $T_t^t=T_z^z>0$ and $T_r^r=T_\theta^\theta=0$ describe classical extended cosmic strings, as described for example in Linet \cite{lin}. 

Note also that in our case, we can solve explicitly for $f$ and $B$ as the equations
\begin{eqnarray}\label{517}
f'
&=& \pm\frac{B}{B_0}\frac{n}{r}\sin f~\text{and}~\left(\frac{B}{B_0}\right)'=\frac{\varepsilon}{\lambda^2}\left(\frac{B}{B_0}\right)^3\frac{n^2}{r}\sin^2f\nonumber\\
\end{eqnarray}
can be combined to give
\begin{eqnarray}\label{518}
f''
&=& \frac{\varepsilon}{\lambda^2}rf'^3 -\frac{f'}{r} +f'^2\cot f
\end{eqnarray}
After multiplying through by $(rf'^2)^{-1}\sin f$, we have
\begin{eqnarray}\label{519}
\frac{\sin f}{rf'^2}f''
&=& \frac{\varepsilon}{\lambda^2}f'\sin f -\frac{\sin f}{r^2 f'} +\frac{\cos f}{r}
\end{eqnarray}
or equivalently
\begin{eqnarray}\label{520}
\frac{d}{dr}\left(\frac{\sin f}{rf'}\right) 
&=& \frac{\varepsilon}{\lambda^2}\frac{d}{dr}(\cos f)
\end{eqnarray}
and so
\begin{eqnarray}\label{521}
\frac{\sin f}{rf'} -\frac{\varepsilon}{\lambda^2}\cos f = K_1
\end{eqnarray}
for some integration constant $K_1$. 

Given that $f(0)=\pi$ and $B(0)=B_0$, the equation
\begin{eqnarray}\label{522}
f'
&=& \pm\frac{B}{B_0}\frac{n}{r}\sin f
\end{eqnarray}
with the negative sign chosen gives $(rf')^{-1}\sin f\rightarrow -n^{-1}$ as $r\rightarrow 0$, and so the integration constant $K_1$ is
\begin{eqnarray}\label{523}
K_1
&=& \lim_{r\rightarrow 0}\left(\frac{\sin f}{rf'} -\frac{\varepsilon}{\lambda^2}\cos f\right)=\frac{\varepsilon}{\lambda^2} -\frac{1}{n}
\end{eqnarray}

Multiplying the equation
\begin{eqnarray}\label{524}
\frac{\sin f}{rf'} -\frac{\varepsilon}{\lambda^2}\cos f =\frac{\varepsilon}{\lambda^2} -\frac{1}{n}
\end{eqnarray}
by $f'/\sin f$ then gives
\begin{eqnarray}\label{525}
\frac{1}{r} -\frac{\varepsilon}{\lambda^2}f'\cot f =\left(\frac{\varepsilon}{\lambda^2} -\frac{1}{n}\right)\frac{f'}{\sin f}
\end{eqnarray}
which if integrated with respect to $r$ becomes
\begin{eqnarray}\label{526}
\ln r -\frac{\varepsilon}{\lambda^2}\ln(\sin f) =-\left(\frac{\varepsilon}{\lambda^2} -\frac{1}{n}\right)\ln\left(\frac{1+\cos f}{\sin f}\right) +K_2 \nonumber\\
\end{eqnarray}
where $K_2$ is a second integration constant. 

Now, the asymptotic relation $(rf')^{-1}\sin f\rightarrow -n^{-1}$ as $r\rightarrow 0$ combined with the initial condition $f(0)=\pi$ solves to give
\begin{eqnarray}\label{527}
f\approx \pi -\alpha r^n
\end{eqnarray}
for small values of $r$, where $\alpha$ is some positive constant (which is equivalent to $-a$ in the models we have been considering so far), and therefore
\begin{eqnarray}\label{528}
\ln r -\frac{\varepsilon}{\lambda^2}\ln (\sin f) \approx \left(1 -\frac{n\varepsilon}{\lambda^2}\right)\ln r -\frac{\varepsilon}{\lambda^2}\ln \alpha
\end{eqnarray}
while
\begin{eqnarray}\label{529}
&&-\left(\frac{\varepsilon}{\lambda^2} -\frac{1}{n}\right)\ln\left(\frac{1+\cos f}{\sin f}\right) +K_2 \nonumber\\
&=&\left(1-\frac{n\varepsilon}{\lambda^2}\right)\left[\ln r +\frac{1}{n}\ln\left(\frac{\alpha}{2}\right)\right] +K_2 \nonumber\\
\end{eqnarray}
Hence,
\begin{eqnarray}\label{530}
K_2
&=& -\frac{1}{n}\ln\alpha +\left(1-\frac{n\varepsilon}{\lambda^2}\right)\frac{1}{n}\ln 2
\end{eqnarray}
and the equation for $f$ becomes
\begin{eqnarray}\label{531}
\ln r -\frac{\varepsilon}{\lambda^2}\ln(\sin f) 
&=& \frac{1}{n}\left(1 -\frac{n\varepsilon}{\lambda^2}\right)\left[\ln\left(\frac{1+\cos f}{\sin f}\right) +\ln 2\right] \nonumber\\
&&-\frac{1}{n}\ln\alpha  
\end{eqnarray}

Multiplying through by $n/(1-n\varepsilon \lambda^{-2})$ and taking exponentials of both sides
\begin{eqnarray}\label{532}
r^{n/(1-n\varepsilon\lambda^{-2})}\sin f^{-n\varepsilon\lambda^{-2}/(1-n\varepsilon\lambda^{-2})}
&=& 2\alpha^{-1/(1-n\varepsilon\lambda^{-2})}\nonumber\\
&&\times \frac{1+\cos f}{\sin f}
\end{eqnarray}
After a little rearrangement, this equation reads
\begin{eqnarray}\label{533}
(\alpha r^n)^{1/(1-n\varepsilon\lambda^{-2})}
&=& 2(1+\cos f)\sin f^{-(1-2n\varepsilon\lambda^{-2})/(1-n\varepsilon\lambda^{-2})}\nonumber\\
\end{eqnarray}

To solve for $B$, note that
\begin{eqnarray}\label{534}
B
&=& -B_0\frac{rf'}{n~\sin f} =\frac{B_0}{1-\frac{n\varepsilon}{\lambda^2}(1+\cos f)}
\end{eqnarray}
as
\begin{eqnarray}\label{535}
\frac{\sin f}{rf'} &=& -\frac{1}{n}\left(1-\frac{n\varepsilon}{\lambda^2}\right) +\frac{\varepsilon}{\lambda^2}\cos f
\end{eqnarray}
from above. 

In particular, the equation for $f$ indicates that $\sin f\rightarrow 0$ and $\cos f\rightarrow 1$ as $r\rightarrow\infty$, so $f\rightarrow 0$ as expected. Hence,
\begin{eqnarray}\label{536}
\lim_{r\rightarrow\infty}B(r) = \frac{B_0}{1-\frac{2n\varepsilon}{\lambda^2}}
\end{eqnarray}
and since we have been requiring the coordinate $r$ to be the physical radius at large distances from the string, we should have $\lim_{r\rightarrow \infty}B(r)=1$. So
\begin{eqnarray}\label{537}
B_0
&=& 1-\frac{2n\varepsilon}{\lambda^2}
\end{eqnarray}

It should be noted that the solutions constructed here, which include the self-gravity of the string, are scale-free just as the non-gravitating $(\varepsilon=0)$ solutions are scale-free. That is to say, the equation for $f$ can be written in the form
\begin{eqnarray}\label{538}
\bar{r}^{~n/(1-n\varepsilon\lambda^{-2})}
&=&
2(1+\cos f)\sin f^{-(1-2n\varepsilon\lambda^{-2})/(1-n\varepsilon\lambda^{-2})}\nonumber\\
\end{eqnarray}
where $\overline{r}=\alpha^{1/n}r$. So (if $n$ and $\varepsilon/\lambda^2$ are fixed) $f$ is a function of $\overline{r}$ alone, and we can make the string as thin or thick as we like by changing the value of $\alpha$.

Moreover, the mass per unit length $\mu$ of the string is
\begin{eqnarray}\label{539}
\mu
&=& \frac{2\pi n^2}{\lambda^2}\int_0^\infty\frac{\sin^2f}{1-\frac{n\varepsilon}{\lambda^2}(1+\cos f)}\frac{dr}{r}
\end{eqnarray}
and since $dr/r=d\overline{r}/\overline{r}$ the mass per unit length is independent of the value chosen for $\alpha$. 

In fact, we can evaluate $\mu$ explicitly by noting that, from the equation relating $f$ and $\overline{r}$
\begin{eqnarray}\label{540}
\frac{n}{1-\frac{n\varepsilon}{\lambda^2}}\frac{d\bar{r}}{\bar{r}}
&=& -\left(1-\frac{n\varepsilon}{\lambda^2}\right)^{-1}\frac{1-\frac{n\varepsilon}{\lambda^2}(1+\cos f)}{\sin f}df \nonumber\\
\end{eqnarray}
and so
\begin{eqnarray}\label{541}
\mu
&=& \frac{4\pi n}{\lambda^2}
\end{eqnarray}

Plots of $f$ and $B$ as functions of $r$ in the case $n=1$ and $\varepsilon/\lambda^2=0.1$ (and so $B_0=0.8$) when $\alpha=1/2$, 1 and 2 are given below
\begin{center}
\includegraphics[scale=0.45]{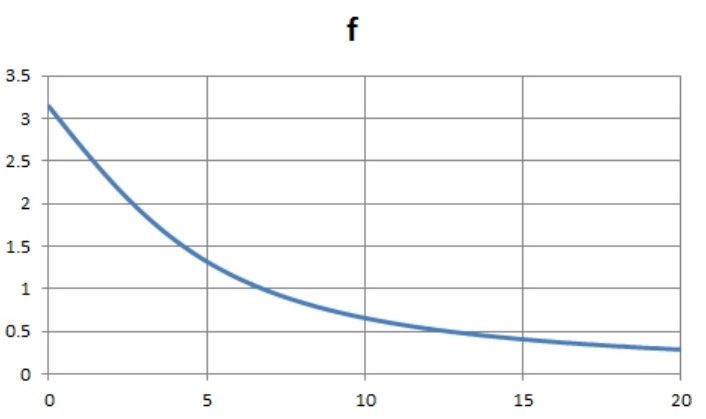}\\
\textbf{Fig.2} Solutions with $K_s=0$ and no-twist.\\
Field function $f$ for $0\leq r\leq 20$ with $\alpha=1/2$
\end{center}

\begin{center}
\includegraphics[scale=0.45]{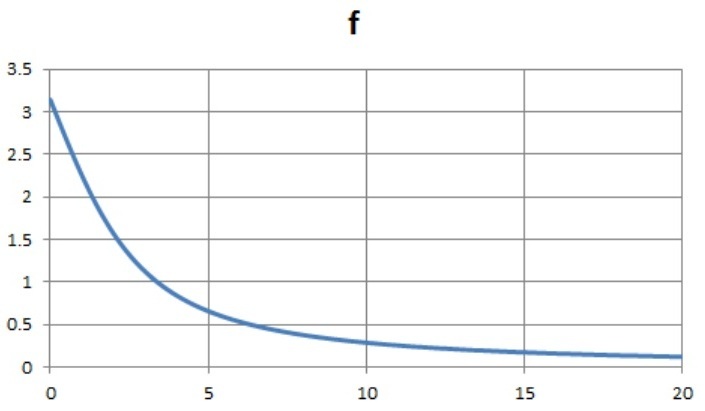}\\
\textbf{Fig.3} Solutions with $K_s=0$ and no-twist.\\
Field function $f$ for $0\leq r\leq 20$ with $\alpha=1$
\end{center}

\begin{center}
\includegraphics[scale=0.45]{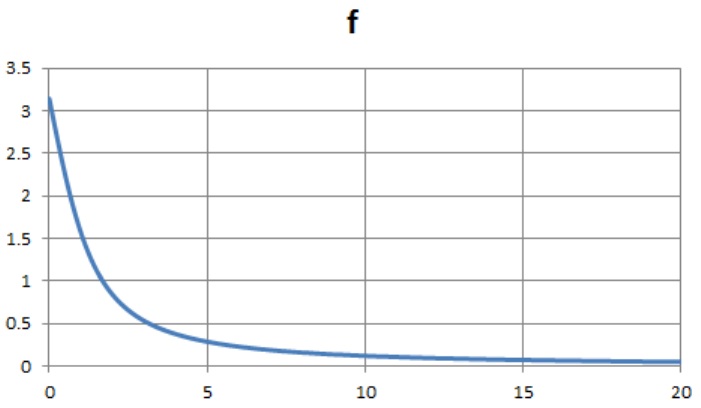}\\
\textbf{Fig.4} Solutions with $K_s=0$ and no-twist.\\
Field function $f$ for $0\leq r\leq 20$ with $\alpha=2$
\end{center}

\begin{center}
\includegraphics[scale=0.45]{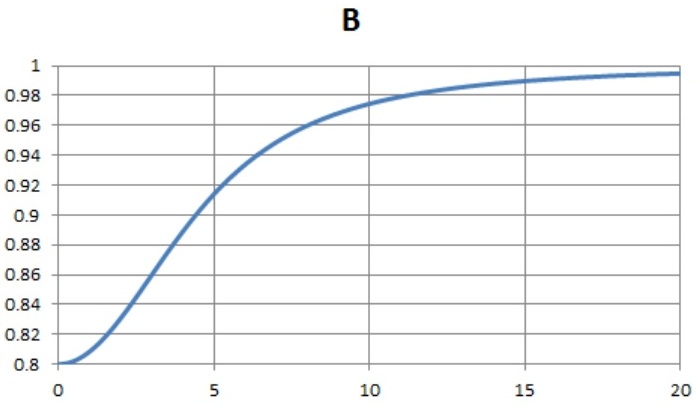}\\
\textbf{Fig.5} Metric function $B$ for $0\leq r\leq 20$ with $\alpha=1/2$
\end{center}

\begin{center}
\includegraphics[scale=0.45]{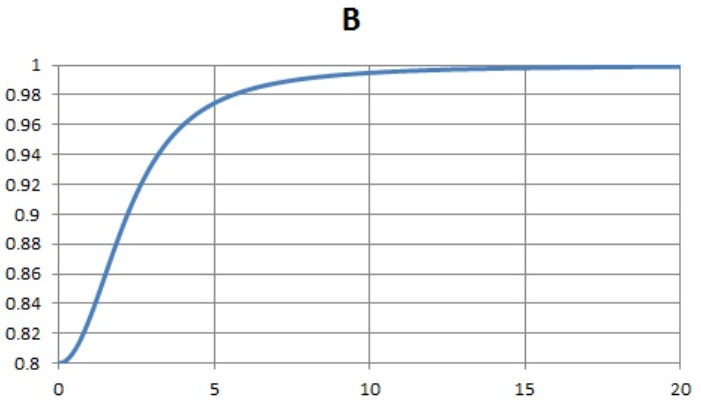}\\
\textbf{Fig.6} Metric function $B$ for $0\leq r\leq 20$ with $\alpha=1$
\end{center}

\begin{center}
\includegraphics[scale=0.45]{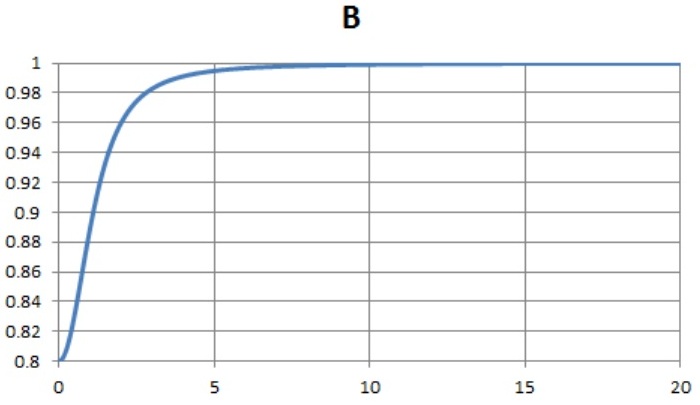}\\
\textbf{Fig.7} Metric function $B$ for $0\leq r\leq 20$ with $\alpha=2$
\end{center}

\begin{center}
\includegraphics[scale=0.45]{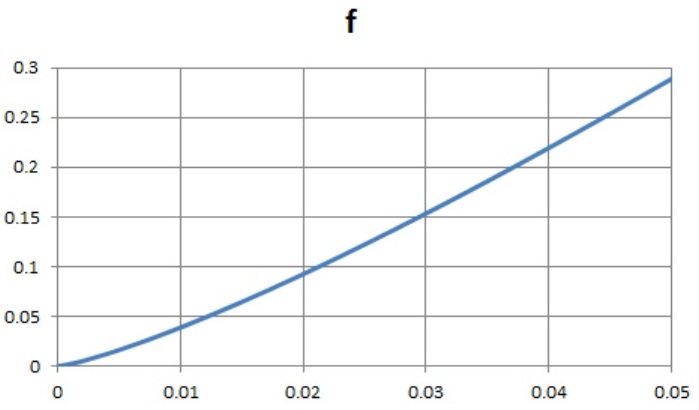}\\
\textbf{Fig.8} Field function $f$ for $0\leq \frac{1}{r}\leq 0.5$ with $\alpha=1/2$
\end{center}

\begin{center}
\includegraphics[scale=0.45]{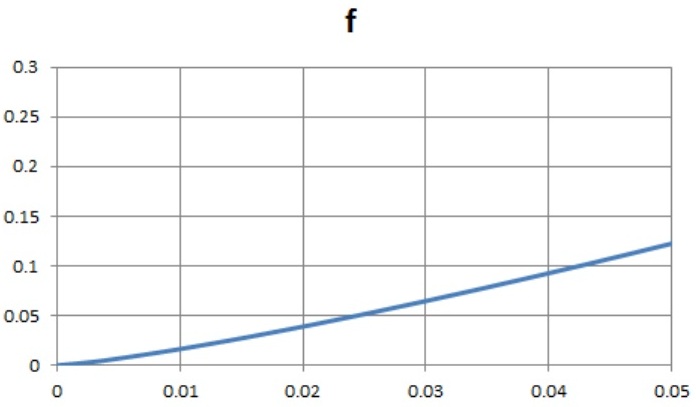}\\
\textbf{Fig.9} Field function $f$ for $0\leq \frac{1}{r}\leq 0.5$ with $\alpha=1$
\end{center}

\begin{center}
\includegraphics[scale=0.45]{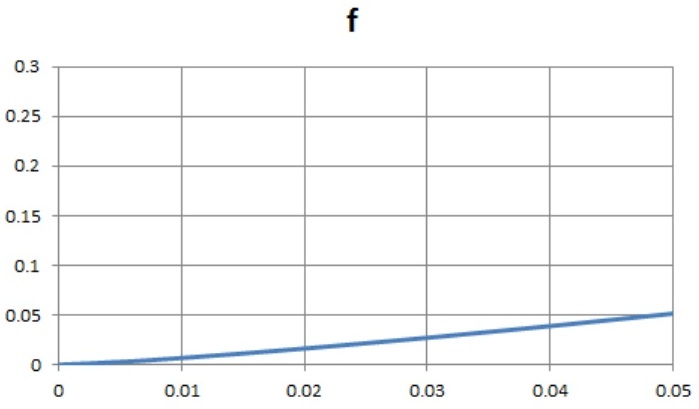}\\
\textbf{Fig.10} Field function $f$ for $0\leq \frac{1}{r}\leq 0.5$ with $\alpha=2$
\end{center}

\begin{center}
\includegraphics[scale=0.45]{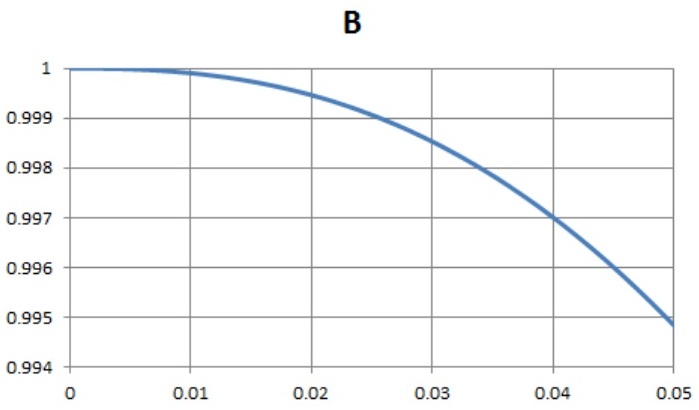}\\
\textbf{Fig.11} Metric function $B$ for $0\leq \frac{1}{r}\leq 0.5$ with $\alpha=1/2$
\end{center}

\begin{center}
\includegraphics[scale=0.45]{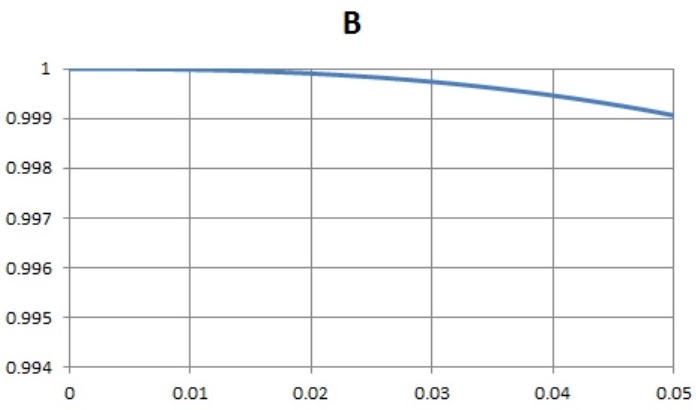}\\
\textbf{Fig.12} Metric function $B$ for $0\leq \frac{1}{r}\leq 0.5$ with $\alpha=1$
\end{center}

\begin{center}
\includegraphics[scale=0.45]{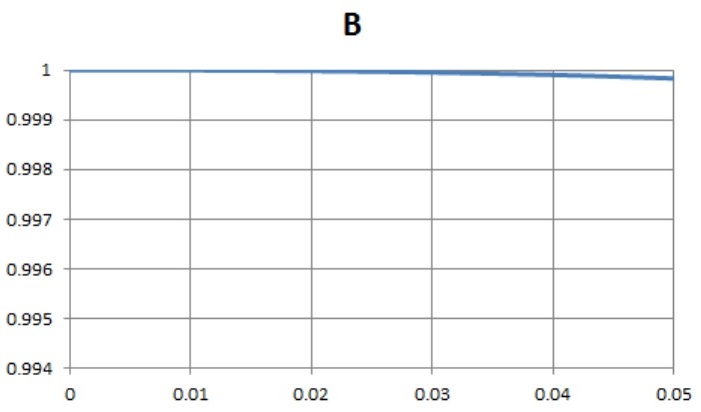}\\
\textbf{Fig.13} Metric function $B$ for $0\leq \frac{1}{r}\leq 0.5$ with $\alpha=2$
\end{center}

\subsection{Twisting-case ($mk\neq 0,~\omega \neq 0$)}
If we assume that $mk\neq 0$ and $\omega\neq 0$ then we still have $A\equiv 1$ but, it is not true that $C=B_0r$ in general. However, there is a special class of exact solutions which have
\begin{eqnarray}\label{542}
C &=& B_0r~~~\text{and}~~~\omega = -\frac{mk}{n}~B_0^2r^2
\end{eqnarray}
This can be verified by substituting these two functions directly into the equations for $C''$ and $\omega''$, as
\begin{eqnarray}\label{543}
C''
&=& 0
\end{eqnarray}
and
\begin{eqnarray}\label{544}
\omega''
&=& -2n^{-1}mkB_0^2
\end{eqnarray}

The two remaining field equations, for $f'$ and $B'$, now read
\begin{eqnarray}\label{545}
B'
&=& \frac{\varepsilon}{\lambda^2}\frac{B^3}{B_0^2}\frac{n^2}{r}\sin^2f
\end{eqnarray}
and
\begin{eqnarray}\label{546}
f'
&=& \pm\frac{B}{B_0}\frac{n}{r}\sin f
\end{eqnarray}
These are exactly the same equations that were generated previously, in the twist free case. So, the solutions are
\begin{eqnarray}\label{547}
B
&=& \frac{B_0}{1-\frac{n\varepsilon}{\lambda^2}(1+\cos f)}
\end{eqnarray}
and
\begin{eqnarray}\label{548}
2(1+\cos f)\sin f^{-(1-2n\varepsilon \lambda^{-2})/(1-n\varepsilon \lambda^{-2})} =\bar{r}^{~n/(1-n\varepsilon\lambda^{-2})}
\end{eqnarray}
where $\overline{r}=\alpha^{1/n}r$ for some positive constant $\alpha$, and
\begin{eqnarray}\label{549}
B_0
&=& 1-\frac{2n\varepsilon}{\lambda^2}
\end{eqnarray}

Also, because 
\begin{eqnarray}\label{550}
Nr^2
&=& -\frac{n^2}{B_0^2}
\end{eqnarray}
as before, the components of the stress-energy tensor $T_{\mu\nu}$ are
\begin{eqnarray}\label{551}
T_{tt}
&=& \frac{1}{\lambda^2}\frac{n^2}{B_0^2r^2}\sin^2f
\end{eqnarray}
\begin{eqnarray}\label{552}
T_{rr}
&=& 0
\end{eqnarray}
\begin{eqnarray}\label{553}
T_{\theta\theta}
&=& 0
\end{eqnarray}
\begin{eqnarray}\label{554}
T_{\theta z}
&=& 0
\end{eqnarray}
and
\begin{eqnarray}\label{555}
T_{zz}
&=& -\frac{1}{\lambda^2}\frac{n^2}{B_0^4r^2}\sin^2f 
\end{eqnarray}

The line element in this case is
\begin{eqnarray}\label{556}
ds^2
&=& dt^2 - B(r)^2dr^2 -B_0^2r^2d\theta^2 -2n^{-1}mkB_0^2r^2~d\theta~dz \nonumber\\
&&-~\frac{1+n^{-2}m^2k^2B_0^4r^2}{B_0^2}~dz^2 
\end{eqnarray}
and in particular
\begin{eqnarray*}
g^{\theta\theta}
&=& -\frac{1+n^{-2}m^2k^2B_0^4r^2}{B_0^2r^2},~g^{\theta z}=n^{-1}mkB_0^2
\end{eqnarray*}
and
\begin{eqnarray}\label{557}
g^{zz}=-B_0^2
\end{eqnarray}
So
\begin{eqnarray}\label{558}
T_t^t
&=& \frac{1}{\lambda^2}\frac{n^2}{B_0^2r^2}\sin^2f = T_z^z
\end{eqnarray}
exactly as in the twist-free case.

Although the line element appears to be more complicated than in the twist-free case, it is evident that the stress-energy content of the space-time is exactly the same. This is because the twisting line element can be written as
\begin{eqnarray}\label{559}
ds^2
&=& dt^2 -B(r)^2dr^2 -B_0^2r^2(d\theta +n^{-1}mk~dz)^2 \nonumber\\
&&-~B_0^{-2}dz^2
\end{eqnarray}
and so it is just the twist-free line element with $\theta$ replaced by $\theta'=\theta-n^{-1}mkz$ as the angular coordinate. Hence, it is not a different solution at all. [We have searched extensively for a numerical solution with $K_s=0$ and $mk\neq0$ that is different from the solution described here, but we have not been able to find one.]

\section{Discussions and Conclusions}
In order to obtain solutions for the Einstein field equations, first we try to find solutions without Skyrme term ($K_s=0$) for non-twisting and twisting cases.

A solution that is regular on the axis $r=0$ will have $A=A'=0$ there, and so $A''=0$ at $r=0$ as well. This means that $A$ will be constant everywhere, and can always be rescaled so that $A\equiv 1$. 

Although the line element appears to be more complicated than in the twist-free case, it is evident that the stress-energy content of the space-time is exactly the same. This is because the twisting line element can be written as
\begin{eqnarray}\label{559}
ds^2
&=& dt^2 -B(r)^2dr^2 -B_0^2r^2(d\theta +n^{-1}mk~dz)^2 \nonumber\\
&& -B_0^{-2}dz^2
\end{eqnarray}
and so it is just the twist-free line element with $\theta$ replaced by $\theta'=\theta-n^{-1}mkz$ as the angular coordinate. Hence, it is not a different solution at all. [We have searched extensively for a numerical solution with $K_s=0$ and $mk\neq0$ that is different from the solution described here, but we have not been able to find one.]

\section{Acknowledgment}
MH thank to UBD GRS Scholarships for supporting this research. 
\\

\end{document}